\documentstyle[12pt,psfig]{article}

\newcommand{\beq}{\begin{equation}}
\newcommand{\eeq}{\end{equation}}
\newcommand{\beqa}{\begin{eqnarray}}
\newcommand{\eeqa}{\end{eqnarray}}

\newcommand{\Eq}[1]{(\ref{eq:#1})}
\newcommand{\Fig}[1]{Fig.~\ref{fg:#1}}

\newcommand{\ct}[1]{\cite{#1}}

\newcommand{\Ham}{{\cal H}}

\newcommand{\deffig}[5]{
\begin{figure}[t]
\begin{minipage}[t]{8cm}
\caption[*]{{\small #5}}
\label{#1}
\end{minipage}
\begin{minipage}[t]{8cm}
\null\ 
\protect{\psfig{figure={#2},height={#3},width={#4}}}
\end{minipage}
\end{figure}
}

\begin{document}

\begin{flushleft}
{\LARGE \bf 
Finite Size Scaling Analysis of Exact Ground States for $\pm J$
Spin Glass Models in Two Dimensions
} \\[5mm]
\end{flushleft}
\null\hspace*{1cm}
\begin{minipage}[t]{15cm}
{\large Naoki~Kawashima$(*)$ and Heiko~Rieger$(**)$ } \\[5mm]
{\it $(*)$ Department of Physics, Toho University, Miyama 2-2-1, 
Funabashi 274, Japan} \\
{\it $(**)$ HLRZ c/o Forschungszentrum J\"ulich, 52425 J\"ulich, Germany} 
\\[5mm]
\today \\[5mm]
PACS.\ 75.10.Nr, 75.50.Lk, 05.50.+q 
--- Spin Glass, Random Systems, Ising models \\[8mm]
{\small
{\bf Abstract. ---}
With the help of {\it exact} ground states obtained by a polynomial
algorithm we compute the domain wall energy $\Delta$ at
zero-temperature for the bond-random and the site-random Ising spin
glass model in two dimensions.  We find that in both models the
stability of the ferromagnetic {\it and} the spin glass order ceases
to exist at a {\it unique} concentration $p_c$ for the ferromagnetic
bonds.  In the vicinity of this critical point, the size and
concentration dependency of the first {\it and} second moment of the
domain wall energy are, for both models, described by a {\it common}
finite size scaling form. Moreover, below this concentration the
stiffness exponent turns out to be slightly negative $\theta_S =
-0.056(6)$ indicating the absence of any intermediate spin glass
phase at non-zero temperature.}\\[8mm]
\end{minipage}


Since Edwards and Anderson (EA) proposed the model for spin
glasses, it has been discussed not only among researchers specialized
for the subject, but in a rather wide community of physicists working
on random systems in general.  Perhaps, this is largely because the EA
model is the simplest possible model with short-ranged interaction for
which we might expect a spin glass phase transition similar to that in
the mean-field model, i.e.\ the Sherrington-Kirkpatrick model.
However, a number of fundamental questions failed to be answered
conclusively \ct{Review}.  Even the very existence of the phase
transition in three dimensions was questioned\ct{Parisi} recently.
Whether or not the spin glass phase exists in the presence of a
uniform magnetic field is even less clear \ct{Caracciolo,KawashimaI}.

Another unanswered question is whether a spin glass phase (at non-zero
temperature) can exist in two dimensions with an {\it asymmetric} bond
distribution.  It has been argued \ct{Barahona82,MaynardR,Ozeki90}
that an intermediate spin glass phase might be present in the $p$-$T$
phase diagram between the ferromagnetic phase and the paramagnetic
phase.  In \Fig{PhaseDiagram}, such a $p$-$T$ phase diagram is shown
including the proposed spin-glass transition line represented by the
dash-dotted line.  For the site-random model the evidence for the
existence of a spin-glass phase seems to be even stronger than for the
bond-random model \ct{ShirakuraM95,OzekiN}.

On the other hand, in the case of the bond-random $\pm J$ model with
$p=1/2$, arguments for the absence of a spin glass phase in two
dimensions were mainly based on results from Monte Carlo simulations
\ct{BhattYoung,MorgensternB} and the estimates of the domain wall
energy \ct{BrayMoore,CieplakB,Ozeki90}. The data from Monte Carlo
simulations, however, are not available at very low temperature, e.g.,
below $T=0.4J$ in Bhatt and Young's simulation \ct{BhattYoung}, which
naturally made it difficult to exclude the possibility of the
transition at a temperature smaller than $0.4J$.  Furthermore it is
not clear, whether the ``stiffness'' exponent
$\theta_S$ is really negative because the data in
Cieplak and Banavar's paper\ct{CieplakB} clearly show a systematic
positive curvature in a log-log plot of the domain wall energy versus
system size for systems without vacancy. 
More recently results of Monte Carlo simulation at lower temperatures
have been reported \ct{ShirakuraM96} indicating a transition at
$T\simeq 0.24J$. In addition, $\theta_S=0$ was suggested \ct{Ozeki90}
based on estimates of the domain-wall energy. These results are
consistent with a finite temperature phase transition for which the
low-temperature phase is only marginally or weakly ordered, meaning
that the two-point spin-spin correlation function decreases
algebraically as a function of the distance.

\deffig{fg:PhaseDiagram}{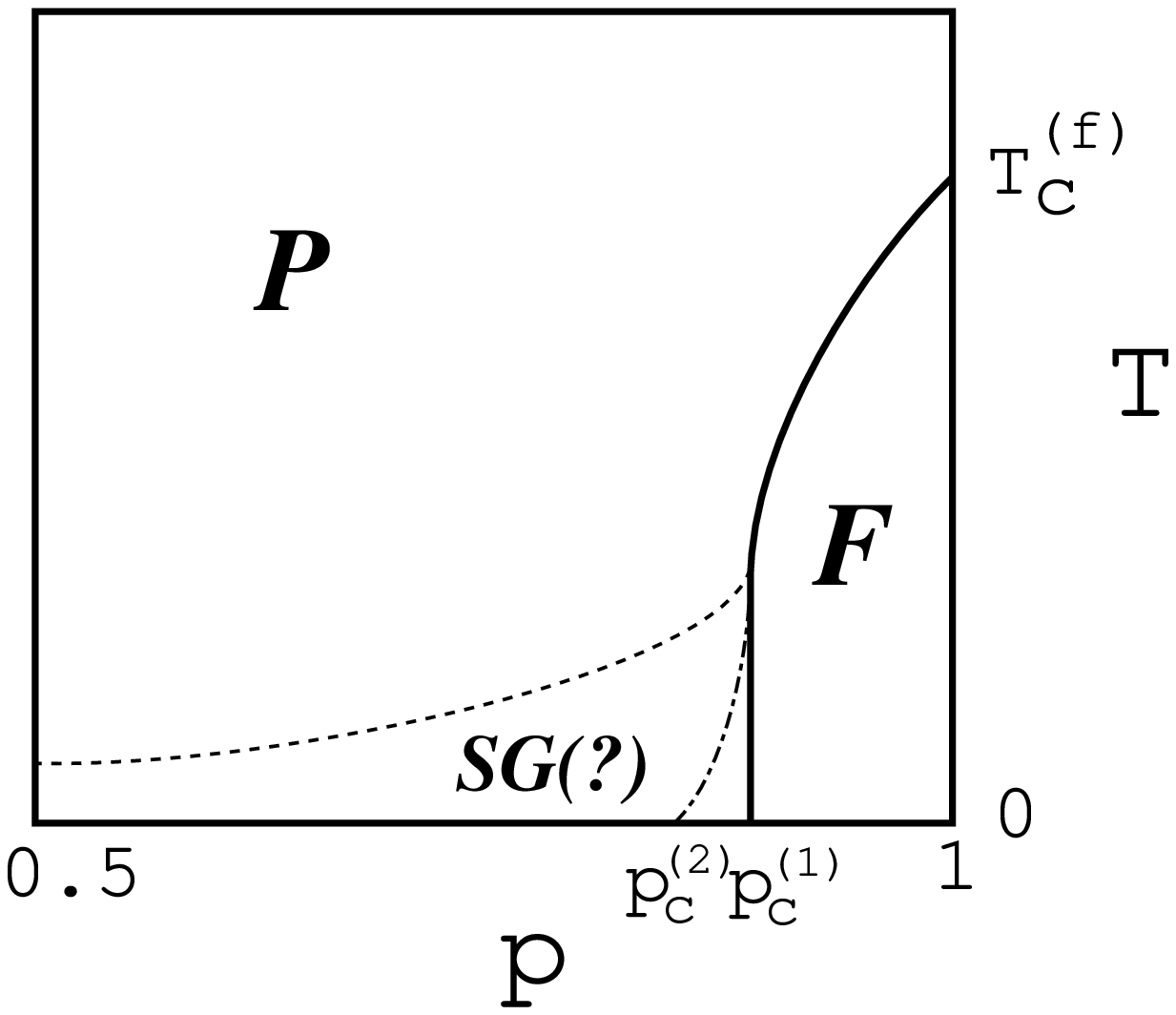}{63mm}{84mm} {The
  schematic phase diagram with the previously proposed spin-glass
  transition lines.  $T_c^{(f)}$ stands for the critical temperature
  of the Ising model on the square lattice.  $P$, $F$, and $SG$ stand
  for the paramagnetic, ferromagnetic and spin glass phases,
  respectively.}


The aim of the present letter is to reinvestigate this issue by
studying the domain wall energy at zero temperature via the
determination of {\it exact} ground states for large system sizes and
huge sample numbers. This can be done very efficiently with the help
of a polynomial algorithm described by Barahona et
al. \ct{Barahona82}, which amounts to finding a
minimum-weight-perfect-matching in a weighted graph with $N=L^2$ nodes
and has computational complexity ${\cal O}(N^3)$. The model that we
consider is the two-dimensional Ising spin glass with binary couplings
defined by the Hamiltonian
\beq
  \Ham \equiv - \sum_{(ij)} J_{ij} S_i S_j
\eeq
where $S_i=\pm1$ are Ising spins, $(ij)$ are nearest neighbor sites on
an $L\times L$-square lattice and the interactions strengths $J_{ij}$
are quenched random variables taking on one of the two values, $+J$
and $-J$.  We consider two different cases: In the bond-random model
all interactions living on the {\it bonds} are independently
distributed with a concentration $p\in[0,1]$ of ferromagnetic bonds
($J_{ij}=+1$). For the site-random model one first generates
independently distributed random variables for all {\it sites},
$\epsilon_i=\pm 1$. The concentration of type-A-sites, i.e.\ those
with $\epsilon=+1$, is $c$, the type-B-sites ($\epsilon=-1$) occur
with probability $1-c$. Then, $J_{ij}$ is set to be $-J$ if and only
if $\epsilon_i = \epsilon_j = -1$, and it is set to be $+J$, otherwise.
In this case the ferromagnetic bond concentration is given by $p
= (2-c)c$.

We calculate the domain wall energy $\Delta$ defined by $\Delta \equiv
E_p - E_a$ where $E_p$ and $E_a$ are the ground state energies with
the periodic and the anti-periodic boundary conditions in the
$x$-direction, respectively.  Free boundary conditions are imposed in
the $y$-direction.  Of crucial importance are the exponents $\rho$ and
$\theta_S$ that characterize the system size dependence of the
moments of the domain wall energy:
\beq
  [\Delta] \propto L^{\rho} \qquad{\rm and}\qquad
  [\Delta^2]^{1/2} \propto L^{\theta_S}
\eeq

A positive value for $\rho$ indicates the stability of a ferromagnetic
ground state even in the presence of thermal fluctuations and thus the
existence of the ferromagnetic long range order at finite temperature
\ct{BrayMoore}. On the other hand, a positive value for the stiffness
exponent, $\theta_S$, with $\rho$ being negative at the same time,
still indicates the stability of the ground state, which possesses now
long range order different from a ferromagnetic one. Thus a positive
$\theta_S$ is interpreted as a sign for a spin glass phase at non-zero
temperature.  We define $p_c^{(1)}$ and $p_c^{(2)}$ as the critical
concentrations of ferromagnetic bonds at which the asymptotic $L$
dependences of $[\Delta]$ and $[\Delta^2]^{1/2}$, respectively, change
from increasing to decreasing, i.e., the concentrations where a
ferromagnetic phase and a spin glass phase, respectively, cease to
exist at finite temperature.


We computed $[\Delta]$ and $[\Delta^2]^{1/2}$ for $L=4,6,8,12,16,24$
and $32$ at various values of $p$ ranging from $0.50$ up to $0.95$.
While the number of bond samples depends on $L$ and $p$, it is 32768
for one of the most time consuming data points, such as the one for
$L=32$ and $p=0.5$.  We hypothesize the following finite size scaling form
for $[\Delta]$
\beq
  [\Delta] L^{\psi_1}
  = f_1( (p-p_c^{(1)}) L^{\phi_1} ). \label{eq:FSSI}
\eeq
The three parameters $p_c^{(1)}$, $\phi_1$ and $\psi_1$ have to be
chosen such that a good data collapse for all data is obtained. To
quantify the ``goodness'' of this fit, we used an appropriate cost
function $S(p_c,\phi,\psi)$ \ct{KawashimaI} whose minimum value
should be close to unity when the fit is statistically acceptable.
When we use all the obtained data points, the best fit is achieved with
$S(p_c,\phi,\psi)\sim 2$, which indicates that there is a
non-negligible systematic error, i.e., a correction to scaling.
Therefore, we have tried a similar analysis on a restricted set of
data, omitting data which are presumably outside the asymptotic
scaling regime, namely, data with $p$ far from $p_c^{(1)}$ and data
for small system sizes. For instance, the goodness of the fit can be
significantly improved to $S=1.12$ by omitting the data for $p=0.95$,
which yields the estimates
\beq
  p_c^{(1)} = 0.896(1),\quad \phi_1=0.77(1),\quad \psi_1=-0.19(2).
  \label{eq:EstimateBondI}
\eeq

\deffig{fg:BondDeltaIFSS}{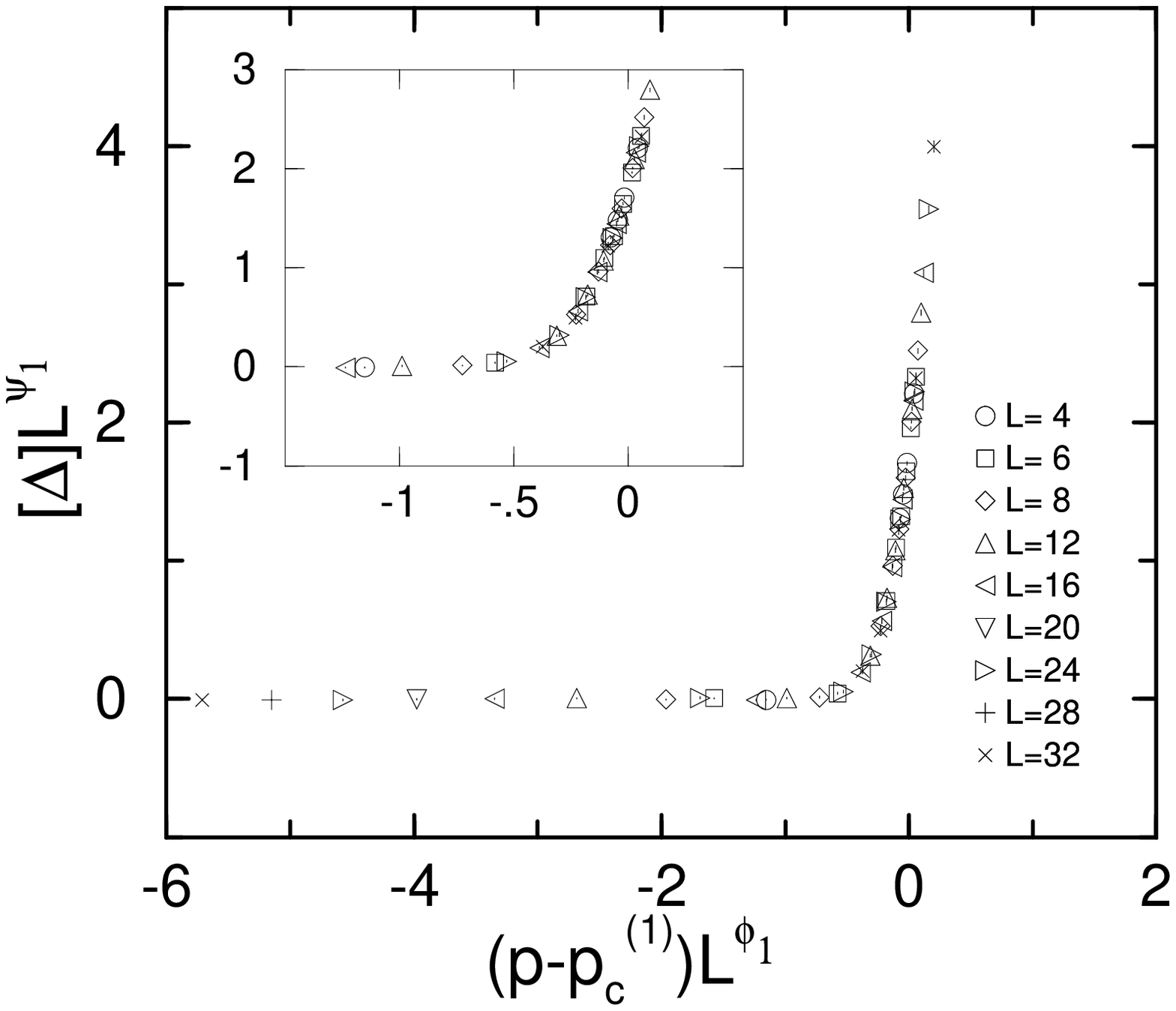}{63mm}{84mm} {The scaling
  plot of $[\Delta]$ for the bond-random model.  $p_c^{(1)} = 0.896$,
  $\phi_1 = 0.77$ and $\psi_1 = -0.19$ are assumed.  The inset is the
  view focused on the region near $p=p_c^{(1)}$.  }

Considering the very small errors accompanying the data points, it is
remarkable that all the data even including those for $L=4$ can be
expressed by the finite size scaling form \Eq{FSSI}.  The resulting
scaling plot is shown in \Fig{BondDeltaIFSS}.  We have confirmed that
other choices of data points producing a value of $S$ close to unity
result in estimates of $p_c^{(1)}$, $\phi_1$, and $\phi_1$
consistent with the estimates quoted above.  The value of
$p_c^{(1)}$ is consistent with most of previous estimate such as
0.88(2) \ct{MorgensternB80}, 0.89(2) \ct{Barahona82} and 0.89(1)
\ct{ONishimori} while inconsistent with 0.885(1) \ct{Ozeki90}.

\deffig{fg:BondDeltaII}{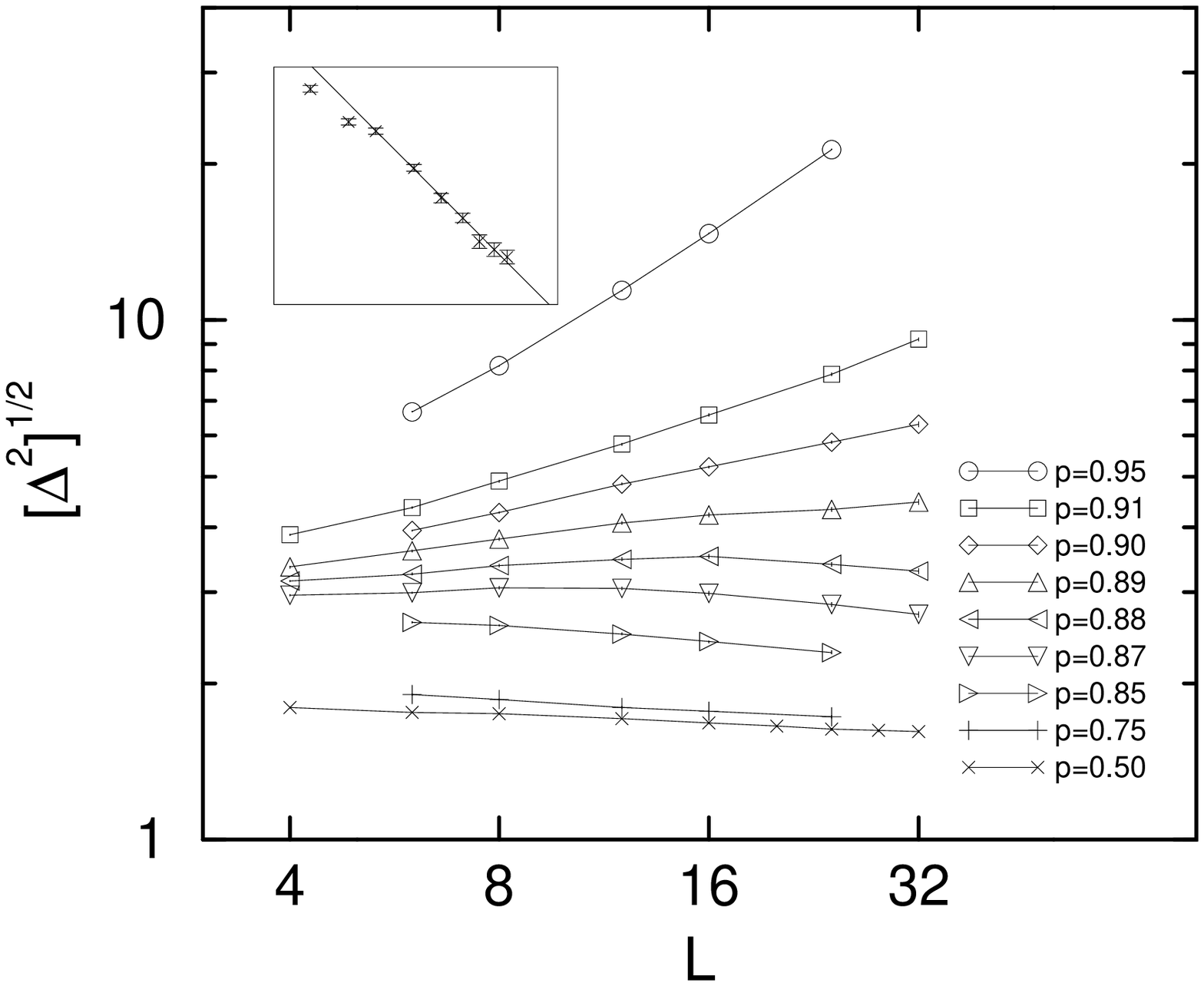}{63mm}{84mm} { The
  domain-wall energy $[\Delta^2]^{1/2}$ of the bond-random model
  plotted against the system size $L$ for various ferromagnetic-bond
  concentration $p$.  The inset is the view focused on the data points
  for $p=1/2$.  The straight line in the inset is obtained by the
  fitting to the data points excluding the two leftmost ones.}

In \Fig{BondDeltaII}, $[\Delta^2]^{1/2}$ is plotted against $L$.  The
lowest curve with crosses, which is almost straight with negative but
very small slope, corresponds to $p=1/2$. In other words, the
domain-wall energy decreases systematically but it does so very
slowly.  The method of least squares using all the data points yields
$\theta_S = -0.052(1)$ whereas all but the first two points (for $L=4$
and $6$) results in $\theta_S = -0.060(2)$.  Therefore, we quote here
the value $\theta_S = -0.056(6)$ as our estimate.

Our results disagree with the suggestion $\theta_S=-0.25$ by Cieplak
and Banavar\ct{CieplakB}.  Considering the size of actual reduction in
$[\Delta^2]^{1/2}$ as $L$ grows from 4 to 32, we cannot rule out the
possibility that the exact value of this exponent is 0, which means
$0<\lim_{L\to\infty}[\Delta]<\infty$. Such a scenario would be
consistent with a suggestion by Ozeki\ct{Ozeki90}.
In this case one has
a marginal situation and we cannot decide whether the long-range
order persists at a low but finite temperature based solely on a
calculation of the stiffness exponent.  We may, however, say that the
low-temperature phase is only weakly ordered even if the phase
transition takes place at a finite temperature.

\deffig{fg:BondDeltaIIFSS}{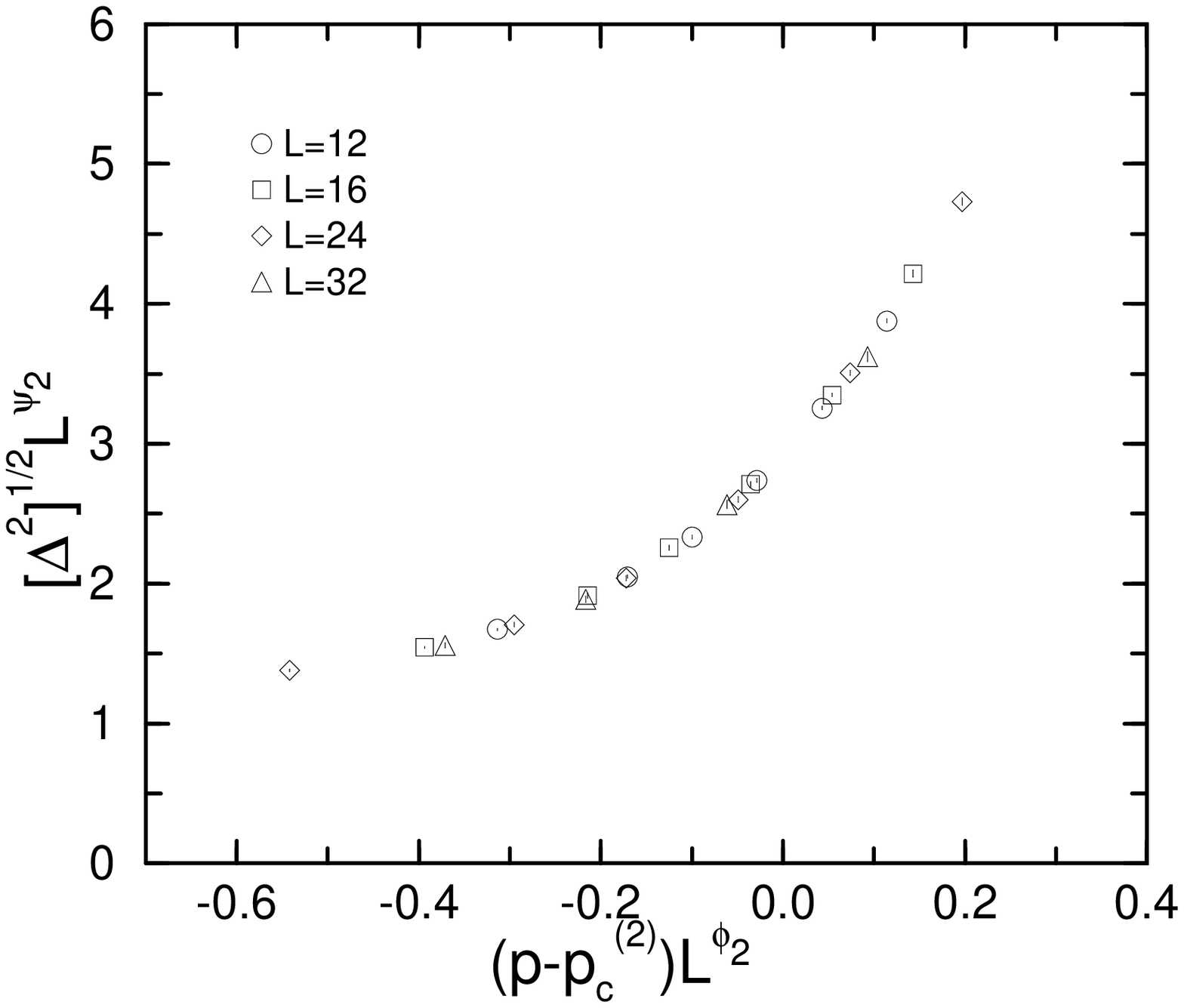}{63mm}{84mm} { The scaling
  plot of $[\Delta^2]^{1/2}$ of the bond-random model.  $p_c^{(2)} =
  0.894$, $\phi_2 = 0.79$ and $\psi_2 = -0.16$ are assumed.  }

Similarly to the above-mentioned procedure employed for $[\Delta]$, we
have tried a finite size scaling analysis for the data of $[\Delta^2]^{1/2}$,
\beq
  [\Delta^2]^{1/2} L^{\psi_2}
  = f_2( (p-p_c^{(2)}) L^{\phi_2} ). \label{eq:FSSII}
\eeq
Now we have to omit more data to get an acceptable fit with the
value of $S$ close to unity, indicating that the correction to scaling
is larger for $[\Delta^2]^{1/2}$ than for $[\Delta]$.  However, a
good data collapse is obtained when we use only data for $L\ge 12$ and
$0.85 \le p \le 0.91$.  The best fit yields
\beq
  p_c^{(2)} = 0.894(2),\quad \phi_2 = 0.79(6),\quad \psi_2 = -0.16(4)
  \label{eq:EstimateBondII}
\eeq
with $S = 0.99$.
The resulting scaling plot is shown in \Fig{BondDeltaIIFSS}.
The present estimate of $p_c^{(2)}$ is larger than 
but marginally consistent with all the previous estimates such as
0.86(2)  \ct{Sadiq},
0.85     \ct{Barahona82} and
0.870    \ct{Ozeki95b},
while it is clearly inconsistent with 0.854(2) \ct{Ozeki90}.

It is remarkable that not only $p_c^{(2)}$ but also $\phi_2$ and
$\psi_2$ agree with the corresponding values in \Eq{EstimateBondI}
within the statistical errors.  While the agreement in $p_c$ already
suggests the absence of the intermediate phase, we consider the
agreement in the critical indices as another evidence for the absence
of the intermediate spin-glass phase, since it is hardly possible
that the first and the second moment of $\Delta$ show the same
critical behavior at different values of $p_c$.


\deffig{fg:SiteDeltaII}{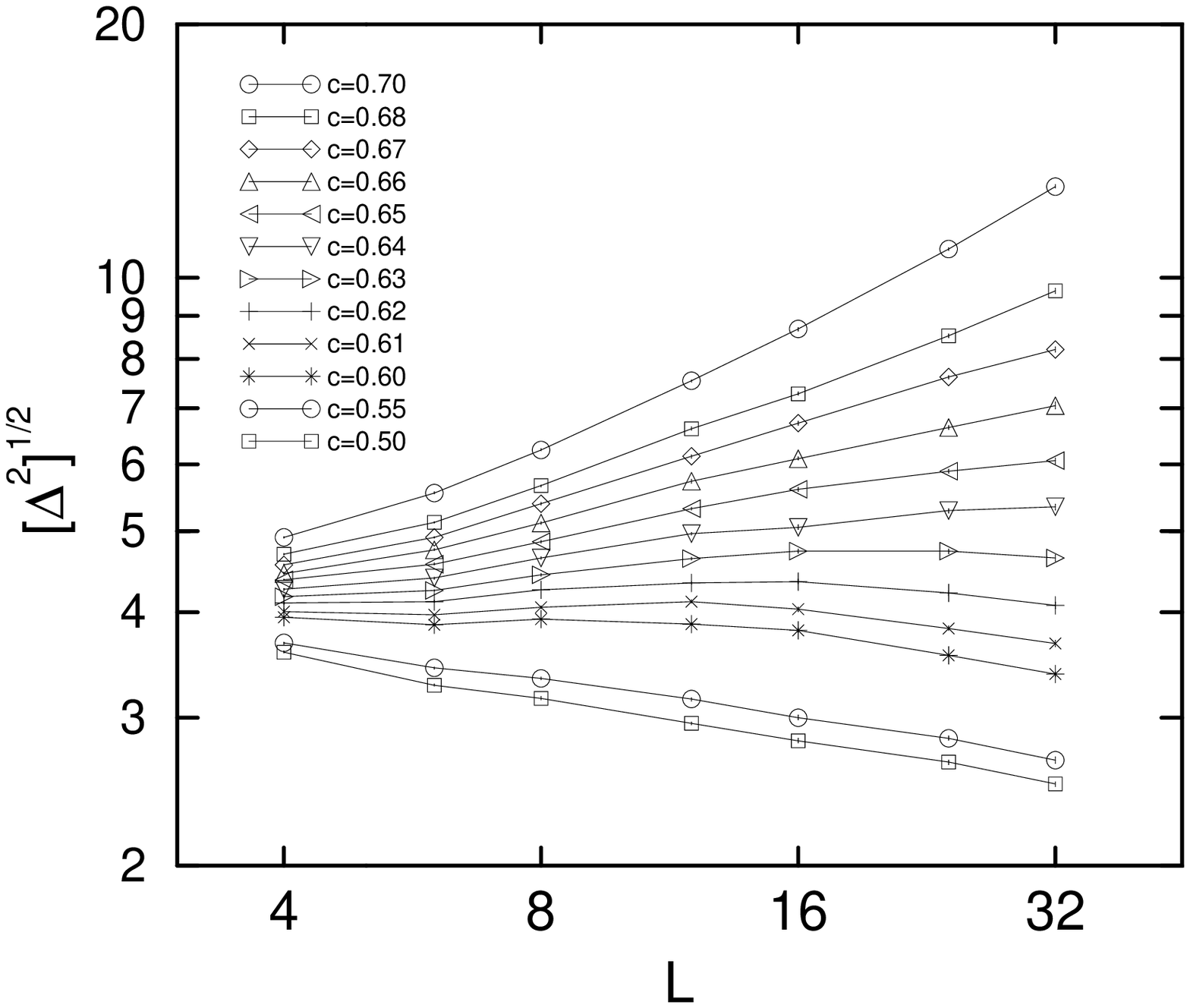}{63mm}{84mm} { The
  domain-wall energy $[\Delta^2]^{1/2}$ of the site-random model
  plotted against the system size $L$ for various `A-site'
  concentration $c$.  }

We now focus on the site-random model.  In \Fig{SiteDeltaII},
$[\Delta^2]^{1/2}$ is plotted as a function of the system size.  A
significant correction to scaling can be seen in \Fig{SiteDeltaII}.
The same remark applies to the first moment, $[\Delta]$.  We have
performed a finite-size-scaling analysis similar to what has been done
for the bond-random model.  As for $[\Delta]$, in order to reduce the
cost function, $S$, down to unity, the smallest system sizes $L=4$ and
$L=6$ have to be excluded from the scaling plot.  The data out of the
range, $0.60 \le c \le 0.68$, are also excluded in the quantitative
estimation of $c_c$ and the indices $\phi_1$ and $\psi_1$.  As for
$[\Delta^2]^{1/2}$, an even stronger correction to scaling is present
as shown in \Fig{SiteDeltaII},
making an additional system size ($L = 8$) unavailable for the
quantitative estimation of $c_c$.  The range of $c$ from which the
data are chosen is again $0.60 \le c \le 0.68$.

The critical concentration, $c_c^{(i)}$, and the critical indices,
$\phi_i$ and $\psi_i$, are defined in a similar fashion to \Eq{FSSI}
and \Eq{FSSII}, resulting in
\beqa
  & & c_c^{(1)} = 0.658(3),\quad \phi_1 = 0.78(2),\quad \psi_1 = -0.18(3) \\
  & & c_c^{(2)} = 0.661(4),\quad \phi_2 = 0.79(2),\quad \psi_2 = -0.23(3)
\eeqa
Obviously $c_c^{(1)} = c_c^{(2)}$ within the error bars, from which
also we here conclude that no intermediate spin glass phase exists. In
addition, again, the critical indices for $[\Delta]$ agree with those
for $[\Delta^2]^{1/2}$ within the statistical errors.
These critical indices agree with those for the bond random model
(\Eq{EstimateBondI} and \Eq{EstimateBondII}), implying that {\it both
models belong to the same universality class}. 

To summarize, we have performed a systematic calculation of the domain
wall energy at zero temperature with systems larger than previous
calculations for both bond-random and site-random models.  We observed
a significant cross-over effect or a correction-to-scaling especially
for the site random model, while no indication for a finite
temperature spin glass phase could be detected.  Some of the previous
evidences for the positive stiffness exponents were based on systems
smaller than those studied in the present paper and therefore may be
attributed to this cross-over effect.  The critical concentration for
the ferromagnetic bonds and critical indices estimated from $[\Delta]$
agree with those from $[\Delta^2]^{1/2}$, again indicating the absence
of an intermediate phase.  Moreover, the critical indices for the
site-random model agree with those for the bond-random model, which
suggests that both models indeed show the same universal critical
behavior and have qualitatively identical features away from $p_c$. We
have also seen that the domain wall energy seems almost independent of
the system size below $p_c$ for the bond-random model at $p=1/2$.  Our
result of a negative stiffness exponent $\theta_S = -0.056(6)$ is
statistically significant, although it is difficult to exclude the
possibility of $\theta_S=0$ because of the very small change in
$[\Delta^2]^{1/2}$ actually observed.

Several other calculations are still in progress.  We have calculated
the spin glass susceptibility at very low temperature down to
$T=0.05J$, and again found no evidence for the phase transition at a
finite temperature. We have also computed the domain wall energy with
a replica boundary condition similar to the one used in Ozeki's work
\ct{Ozeki95b}. These results will be published elsewhere
\ct{KawashimaRDJ}. As a future perspective it might be promising to
perform a similar study of the two-dimensional Ising spin glass model
with next-nearest neighbor interactions and for which indications of a
spin glass transition have been found \cite{campbell}.


This work was mainly done while one of the authors (N.K.)  was at the
supercomputer center HLRZ c/o Forschungszentrum J\"ulich, Germany. He
would like to thank the HLRZ for its hospitality and financial support.

\end{document}